\documentclass[9pt]{article}
\usepackage{authblk}
\usepackage{graphicx}
\usepackage{textcomp}
\usepackage{abstract}
\usepackage[margin=1in]{geometry}

\title{Photonic-chip supercontinuum with tailored spectra
  for precision frequency metrology}

\author[1,*]{David R. Carlson}
\author[1]{Daniel D. Hickstein}
\author[1,2]{Alex Lind}
\author[1]{Judith B. Olson}
\author[1]{Richard W. Fox}
\author[1]{Roger C. Brown}
\author[1]{Andrew D. Ludlow}
\author[3]{Qing Li}
\author[3]{Daron Westly}
\author[1,2]{Holly Leopardi}
\author[1]{Tara M. Fortier}
\author[3]{Kartik Srinivasan}
\author[1,2]{Scott A. Diddams}
\author[1,2]{Scott B. Papp}

\affil[1]{Time and Frequency Division, National Institute of Standards
  and Technology, 325 Broadway, Boulder, CO, 80305}

\affil[2]{Dept. of Physics, University of Colorado, 2000 Colorado Ave, Boulder,
  CO, 80309}

\affil[3]{Center for Nanoscale Science and Technology, National
  Institute of Standards and Technology, 100 Bureau Drive, Gaithersburg, Maryland, 20899}

\affil[*]{Corresponding author: david.carlson@nist.gov}

\date{}

\begin{document}

\maketitle

\begin{abstract}
  Supercontinuum generation using chip-integrated photonic waveguides is a
  powerful approach for spectrally broadening pulsed laser sources with very low
  pulse energies and compact form factors.  When pumped with a mode-locked laser
  frequency comb, these waveguides can coherently expand the comb spectrum to
  more than an octave in bandwidth to enable self-referenced
  stabilization. However, for applications in frequency metrology and precision
  spectroscopy, it is desirable to not only support self-referencing, but also
  to generate low-noise combs with customizable broadband spectra.  In this
  work, we demonstrate dispersion-engineered waveguides based on silicon nitride
  that are designed to meet these goals and enable precision optical metrology
  experiments across large wavelength spans. We perform a clock comparison
  measurement and report a clock-limited relative frequency instability of
  $3.8\times10^{-15}$ at $\tau = 2$~seconds between a 1550~nm cavity-stabilized
  reference laser and NIST's calcium atomic clock laser at 657~nm using a
  two-octave waveguide-supercontinuum comb.
\end{abstract}


\renewcommand{\floatpagefraction}{.8}

\section{Introduction}
Integrated photonic waveguides based on stoichiometric silicon nitride
(Si$_3$N$_4$, henceforth SiN) are a powerful alternative to nonlinear fibers for
generating broadband supercontinuum (SC)~\cite{halir_ultrabroadband_2012,
  chavez_boggio_dispersion_2014, zhao_visible--near-infrared_2015,
  epping_-chip_2015}. Though other materials like
silicon-on-insulator~\cite{hsieh_supercontinuum_2007,
  kuyken_octave-spanning_2015}, silica~\cite{oh_supercontinuum_2014,
  oh_coherent_2017}, chalcogenide glasses~\cite{eggleton_chalcogenide_2011}, and
AlGaAs~\cite{dolgaleva_broadband_2010} are also suitable for chip-based
nonlinear optics, the comparative advantages of SiN are its high nonlinearity,
complementary metal-oxide-semiconductor (CMOS)-compatible fabrication process,
and broad spectral coverage ranging from the visible to the
mid-infrared~\cite{johnson_octave-spanning_2015, epping_-chip_2015}.
Additionally, photonic waveguide devices feature highly tunable dispersion, as
well as high confinement of the light, and offer a potential avenue for
broadening combs with repetition rates \textgreater10~GHz such as low-power microresonator
combs~\cite{kippenberg_microresonator-based_2011, herr_temporal_2013},
electro-optic combs~\cite{kobayashi_optical_1988, ishizawa_octave-spanning_2010,
  supradeepa_bandwidth_2012}, and even some traditional mode-locked
lasers~\cite{bartels_10-ghz_2009}. While SiN waveguides have been demonstrated
to support $f$-2$f$ self-referencing of mode-locked frequency combs with
repetition rates of 1~GHz or less~\cite{mayer_frequency_2015,
  klenner_gigahertz_2016}, applications in frequency metrology and spectroscopy
demand combs that are broadband, frequency stable, and provide user-defined
spectra.

\begin{figure}
  \centering
  \includegraphics[width=0.5\linewidth]{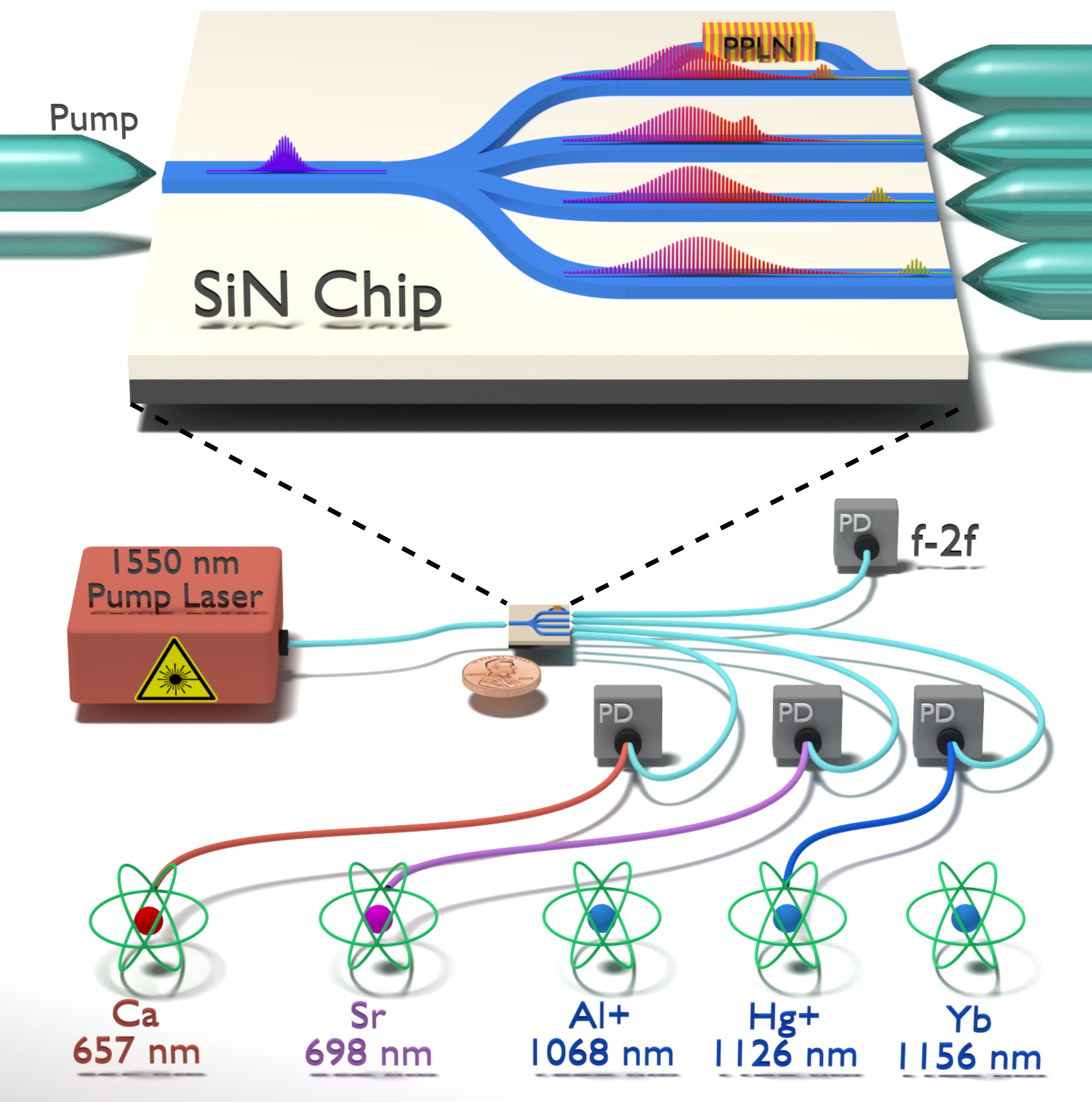}
  \caption{Schematic of proposed all-in-one optical clock comparison chip based
    on the silicon nitride (SiN) waveguide design and experiment presented in
    this work. A single seed comb is split on-chip to direct power to multiple
    waveguides whose dispersion profiles are tuned to produce power at
    wavelengths corresponding to each of the primary optical clock
    standards. The waveguides are out-coupled and delivered in fiber to
    photodetectors (PD) where the SC is overlapped with the appropriate clock
    laser to obtain heterodyne beats. The broad design and fabrication
    capabilities of SiN can also allow integration of frequency doubling
    components such as periodically poled lithium niobate (PPLN) for
    self-referencing~\cite{chang_heterogeneous_2017}.}
  \label{fig:superchip}
\end{figure}

For example, optical-clock comparison measurements require low-noise combs with
frequency bandwidths often spanning hundreds of terahertz. Current techniques
include using octave-spanning Ti:sapphire laser
systems~\cite{fortier_octave-spanning_2006} and multi-branch fiber frequency
combs~\cite{nakajima_multi-branch_2010, hagemann_providing_2013}. However,
chip-integrated devices are now poised to deliver the best features of both
systems. By integrating most wavelength-specific beam paths on a single chip and
eliminating additional amplifiers, waveguide devices can offer reduced
measurement noise and increased sensitivity in a compact form-factor that
promotes portability and low-maintenance operation.


In this work, we demonstrate SiN waveguides designed to support high precision
optical frequency metrology experiments as a key step towards an integrated
all-in-one clock-comparison device shown in Fig.~\ref{fig:superchip}. To show
this, we use the waveguide-generated SC to measure the relative frequency
stability of a 1550~nm cavity-referenced ``clock'' laser versus the
cavity-stabilized 657~nm laser used in the NIST calcium thermal beam atomic
clock~\cite{fox_high_2012, 2015APS..DMP.D1071O}.  The generated SC spectrum
spans from 650~nm to 2.6~\textmu{}m and provides a phase-coherent link between the
1550~nm laser and the calcium clock laser which is over an octave in frequency away
from the pump.  In addition to showing its utility for metrology
experiments, this measurement emphasizes the high temporal coherence,
high-efficiency wavelength conversion, broad spectral bandwidth, and potential
for long-term stability achievable with SC generation in SiN waveguides.

\section{Waveguide Design \& Fabrication}

\begin{figure}
  \centering
  \includegraphics[width=0.5\linewidth]{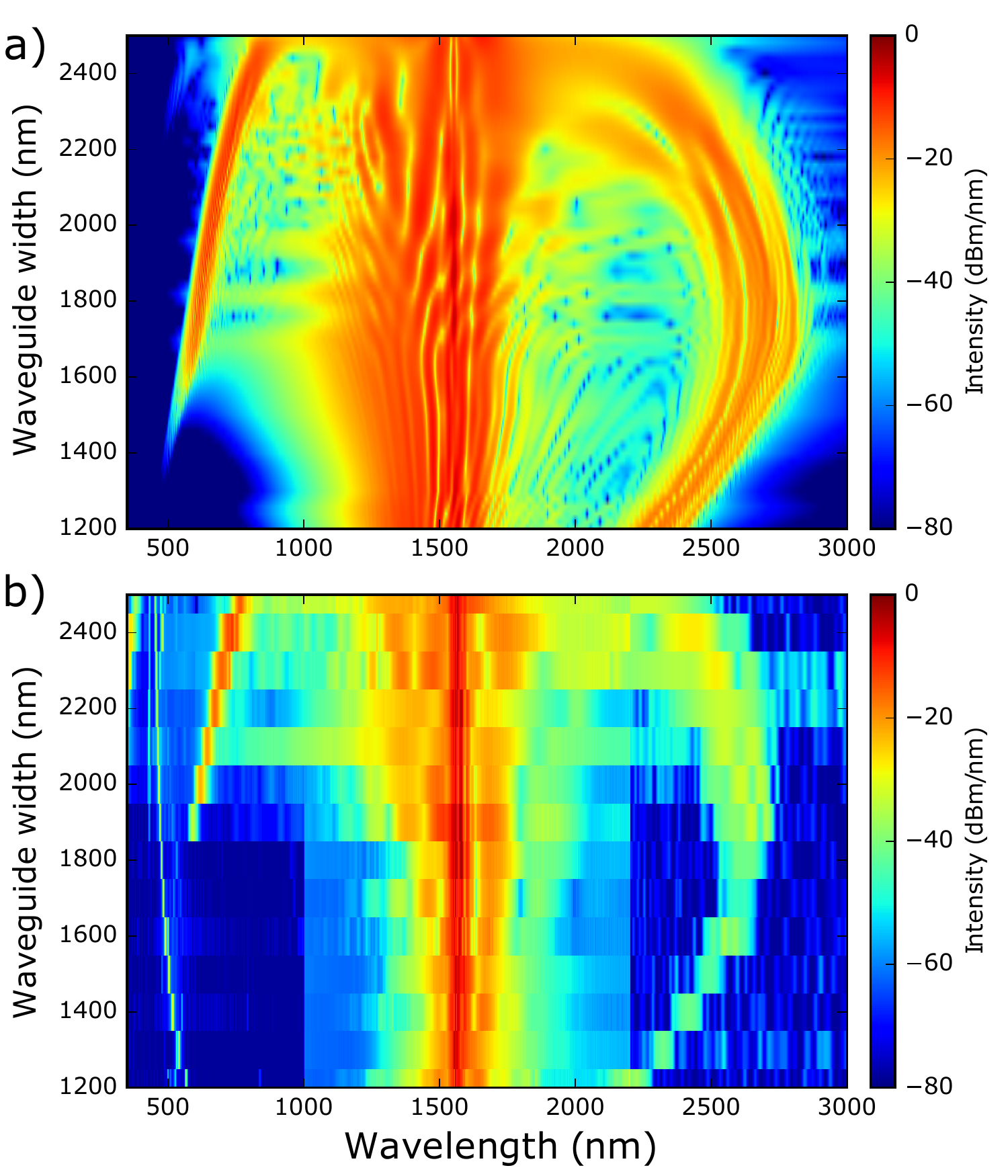}
  \caption{a) Simulated and b) experimental supercontinuum spectra (intra-cavity
    intensity scaled by coupling loss) vs. waveguide width obtained for the 1~cm
    long dispersion-engineered 600~nm air-clad SiN waveguides used in this
    work. A 120~fs pump pulse centered at 1550~nm with total energy of 100~pJ is
    used to seed the waveguide.}
  \label{fig:rainbow}
\end{figure}

To design a suitable waveguide to target a clock-comparison measurement at
657~nm, numerical simulations of the pulse propagation and subsequent spectral
broadening were performed using the generalized nonlinear Schr\"odinger equation
(NLSE) as part of the PyNLO software package ~\cite{ycas_pynlo:_2015,
  hickstein_photonic-chip_2016, amorim_sub-two-cycle_2009}. Included in these
calculations was a chromatic dispersion profile for each waveguide geometry
obtained using a finite difference mode solver implemented in the EMpy software
package~\cite{fallahkhair_vector_2008}.

\begin{figure}
  \centering
  \includegraphics[width=0.5\linewidth]{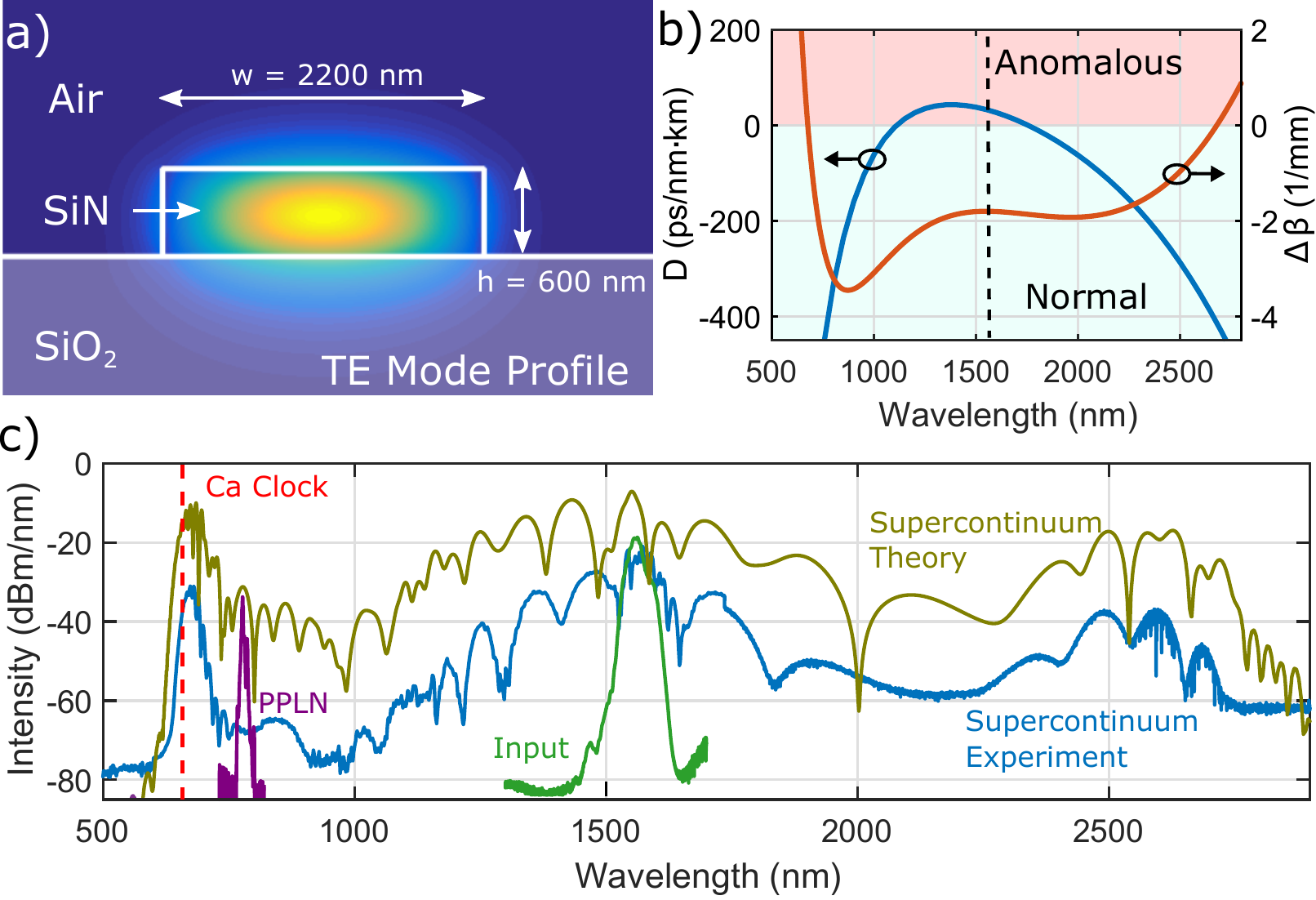}
  \caption{Air-clad SiN waveguide design: a) Waveguide cross-section showing the
    fundamental electric-field TE mode profile for $\lambda = 1550$~nm. b)
    Calculated dispersion profile (left axis) for the waveguide including
    contributions from both the material refractive index and waveguide
    geometry. At the pump wavelength (dotted line), the dispersion is anomalous
    to facilitate soliton compression and broadening. The phase mismatch
    $\Delta\beta$ between the fundamental soliton and a low-amplitude linear
    wave is also shown (right axis).  A dispersive wave occurs in the spectrum
    where $\Delta\beta = 0$. c) Experimental supercontinuum spectrum designed to
    produce a dispersive wave centered at 660~nm (blue).  Also shown are the
    input comb spectrum (green, offset), PPLN spectrum for self-referencing
    (purple, offset), theoretical supercontinuum spectrum (mustard, offset from
    experimental curve due to output coupling loss), and calcium clock
    wavelength at 657~nm (dashed vertical line).}
  \label{fig:spectrum}
\end{figure}

Dispersion engineering is an important feature of SiN waveguides. The high
degree of control over the dispersion arises from the large refractive index
contrast between the SiN core and the lower-index cladding layers to create
strong spatial confinement~\cite{turner_tailored_2006}. As a result, changing
the waveguide geometry can be sufficient to counteract material dispersion
contributions and dramatically alter the output spectrum.  To support soliton
propagation and to achieve the broadest supercontinuum spectrum, it is important
to have anomalous dispersion around the pump
wavelength~\cite{dudley_supercontinuum_2006}. However, dispersive wave
generation providing local spectral enhancement occurs in the normal-dispersion
regime where phase matching is achieved between the fundamental soliton and a
small-amplitude linear wave of different
frequency~\cite{akhmediev_cherenkov_1995, dudley_supercontinuum_2006}. This
phase matching is plotted in Fig.~\ref{fig:spectrum}b as the difference in
wavenumber, $\Delta\beta$, between the soliton and linear wave at different
wavelengths.  The dispersive wave locations, given by $\Delta\beta = 0$, can
thus be precisely tuned through modifications to the waveguide geometry. In
fact, for the metrology experiment described in this work, sufficient tunability
is achieved through tuning the waveguide width alone. Fig.~\ref{fig:rainbow}
highlights the available design space, both through simulation and experiment,
for this spectral tailoring by sweeping the waveguide width while keeping all
other parameters constant.  It was subsequently determined that, for an air-clad
waveguide with a thickness of 600~nm, a waveguide width of 2200~nm would produce
a dispersive wave with the highest amount of optical power near the calcium
clock wavelength. The waveguide geometry and corresponding transverse-electric
(TE) mode profile at 1550~nm is shown in Fig.~\ref{fig:spectrum}a, while the
chromatic dispersion profile is provided in Fig.~\ref{fig:spectrum}b.

With this degree of control over the generated spectrum in the design stage, it
is now possible to simultaneously target other clock wavelengths in parallel
waveguides.  For example, Fig.~\ref{fig:rainbow} shows that to reach the
strontium lattice clock at 698~nm, a waveguide width of 2300~nm should be
chosen.  Optical clock transitions below 600~nm, on the other hand, are commonly
accessed using sub-harmonics of the natural transition wavelengths near 1100~nm
(see Fig.~\ref{fig:superchip}).  While the spectra in Fig.~\ref{fig:rainbow} do
not cover this region well, it is straightforward to extend the dispersive wave
range by widening the waveguide and slightly increasing the thickness of the SiN
layer (see Supplementary Material).  This ``designability'' is a key aspect
allowing the integration of several waveguides onto the same chip as proposed in
Fig.~\ref{fig:superchip} in order to simultaneously target all current optical
clock standards while starting from a fiber-based 1550~nm
source~\cite{hong_optical_2017}.  For the same laser system used here,
implementing this chip would require approximately 150~mW of coupled power.
However, similar fiber lasers with lower repetition rates and shorter pulse
durations can significantly reduce this requirement.

The waveguides used in Figs.~\ref{fig:rainbow}b and \ref{fig:spectrum} are
fabricated by depositing low-pressure chemical vapor deposition (LPCVD)
stoichiometric SiN with a thickness of 600~nm above a 3~\textmu{}m oxide
undercladding layer (SiO$_2$) on a silicon wafer.  The waveguide pattern is then
written to the chip using electron-beam lithography before a final etching step
yields the finished device.  The air-clad waveguides produced here are 1~cm in
length though, to further reduce the pulse energy requirements for dispersive
wave generation, longer waveguides could be used in the future.

\section{Metrology Experiment \& Results}

\begin{figure}
  \centering
  \includegraphics[width=0.5\linewidth]{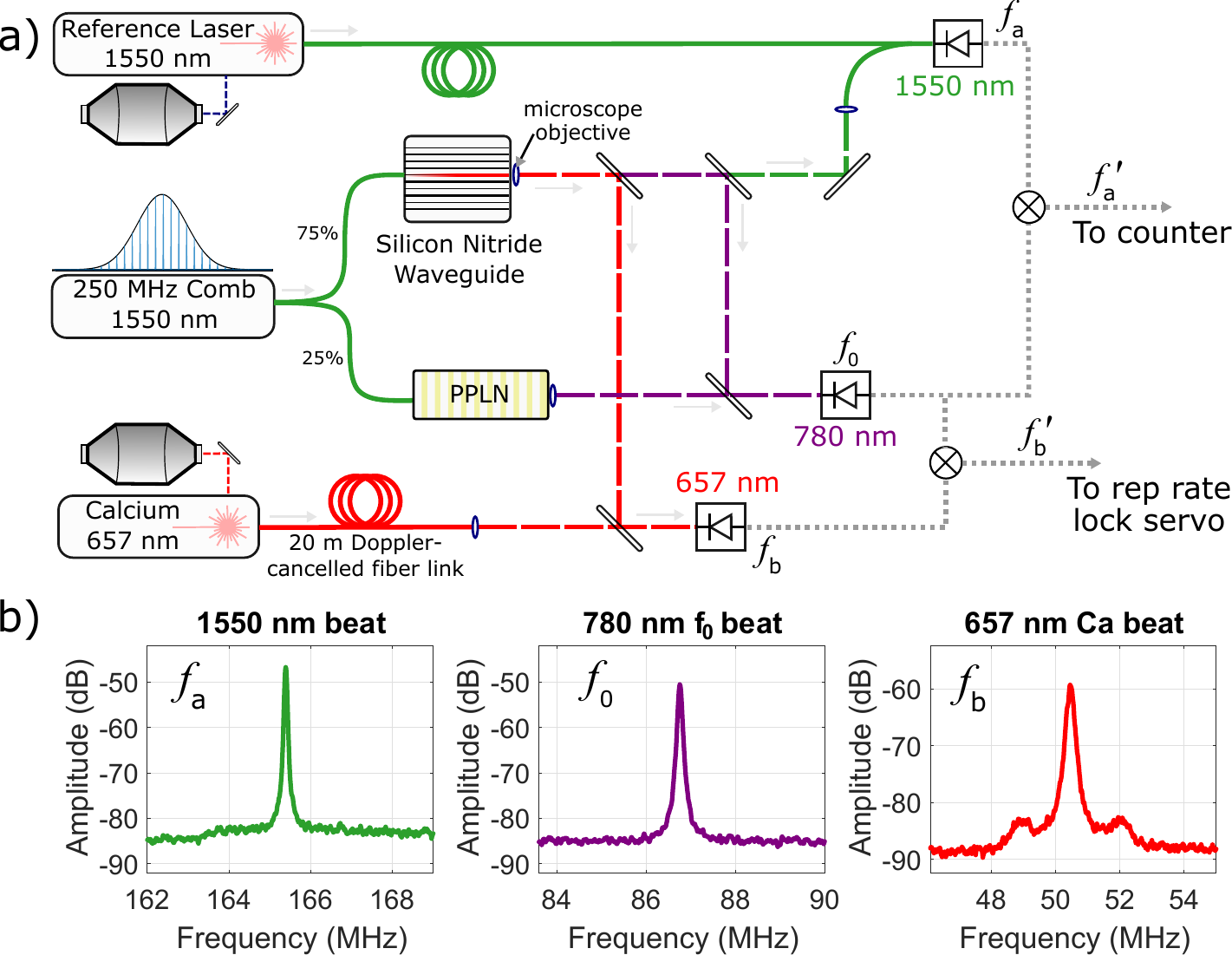}
  \caption{Experimental schematic (fiber path: solid lines, free-space: dashed
    lines, electrical path: dotted gray lines). A mode-locked frequency comb is
    spectrally broadened in a silicon nitride waveguide to produce a spectrum
    spanning two octaves.  Heterodyne beat frequencies $f_a$ and $f_b$, shown in
    b), are obtained between the broadened comb and the cavity-stabilized clock
    lasers while the comb offset frequency, $f_0$, is detected and
    electronically subtracted from both $f_a$ and $f_b$.  The relative stability
    of the optical references is then determined by recording $f'_a = f_a-f_0$
    with a frequency counter while the comb is phase-locked to
    $f'_b = f_b-f_0$.}
  \label{fig:schematic}
\end{figure}

A schematic of the experiment is shown in Fig.~\ref{fig:schematic}. A commercial
1550~nm frequency comb (Menlo Systems M-Comb*) is amplified to an average power
of 300~mW to produce 120~fs pulses at a 250~MHz repetition rate.  A 75\% power
splitter directs light to a lensed fiber for input-coupling to the waveguide
with approximately -7~dB of insertion loss. The remaining 25\% of the amplifier
output is diverted to a 4~cm long waveguide periodically-poled lithium niobate
(PPLN) crystal to generate 780~nm second harmonic light for $f$-2$f$
self-referencing.  Output-coupling from the waveguide is accomplished using a
0.85~NA visible-wavelength microscope objective that collimates the light in
free space.

The experimental SC spectrum showing the calcium-optimized dispersive wave, as
well as both the input and PPLN spectra, is shown in Fig.~\ref{fig:spectrum}c.
Though the doubled PPLN light overlaps with a weak portion of the SC spectrum,
$f$-2$f$ offset detection takes advantage of the coherent addition of many comb
teeth and, as a result, a beat note with a 34~dB signal-to-noise ratio (SNR) at
300 kHz resolution bandwidth (RBW) can still be detected.

Approximately 1~mW of cavity-stabilized calcium clock light is delivered to the
experiment through a Doppler-canceled fiber link~\cite{ma_delivering_1994} that,
when combined with the SC output at a polarizing beam splitter before
photodetection, produces the RF beat $f_b$ shown in
Fig.~\ref{fig:schematic}b. Because the short-wavelength dispersive wave contains
more than 1~nW of optical power per mode, the RF beat readily has >30~dB SNR at
300 kHz RBW, which is an important practical threshold for accurate
stabilization, frequency division, and counting~\cite{hall_stabilizing_1999-1}.
Likewise, another beat note, $f_a$, is obtained from the waveguide output by
combining it with the second ``clock'' laser at 1550~nm.

All three detected RF beats: $f_a$, $f_b$, and $f_0$, are shown in
Fig.~\ref{fig:schematic}b.  The heterodyne signals from single comb lines $f_a$
and $f_b$ are given by
\begin{eqnarray}
f_a &=& n f_r + f_0 - \nu_{1550}\\
f_b &=& m f_r + f_0 - \nu_{657}\nonumber
\end{eqnarray}
where $n$ and $m$ are the comb mode numbers at the clock laser frequencies
$\nu_{1550}$ and $\nu_{657}$, respectively.

After $f$-2$f$ beat detection, RF filtering, and digital frequency division, the
comb offset frequency is electronically subtracted from both $f_a$ and $f_b$
with a double-balanced frequency mixer to obtain offset-free beats $f'_a$ and
$f'_b$~\cite{stenger_ultraprecise_2002}:
\begin{eqnarray}
  \label{eq:fa}
f'_a \quad \equiv\quad f_a - f_0 &=& n f_r - \nu_{1550}\\
f'_b \quad \equiv\quad f_b - f_0 &=& m f_r - \nu_{657}.\nonumber
\end{eqnarray}

\begin{figure}
  \centering
  \includegraphics[width=0.5\linewidth]{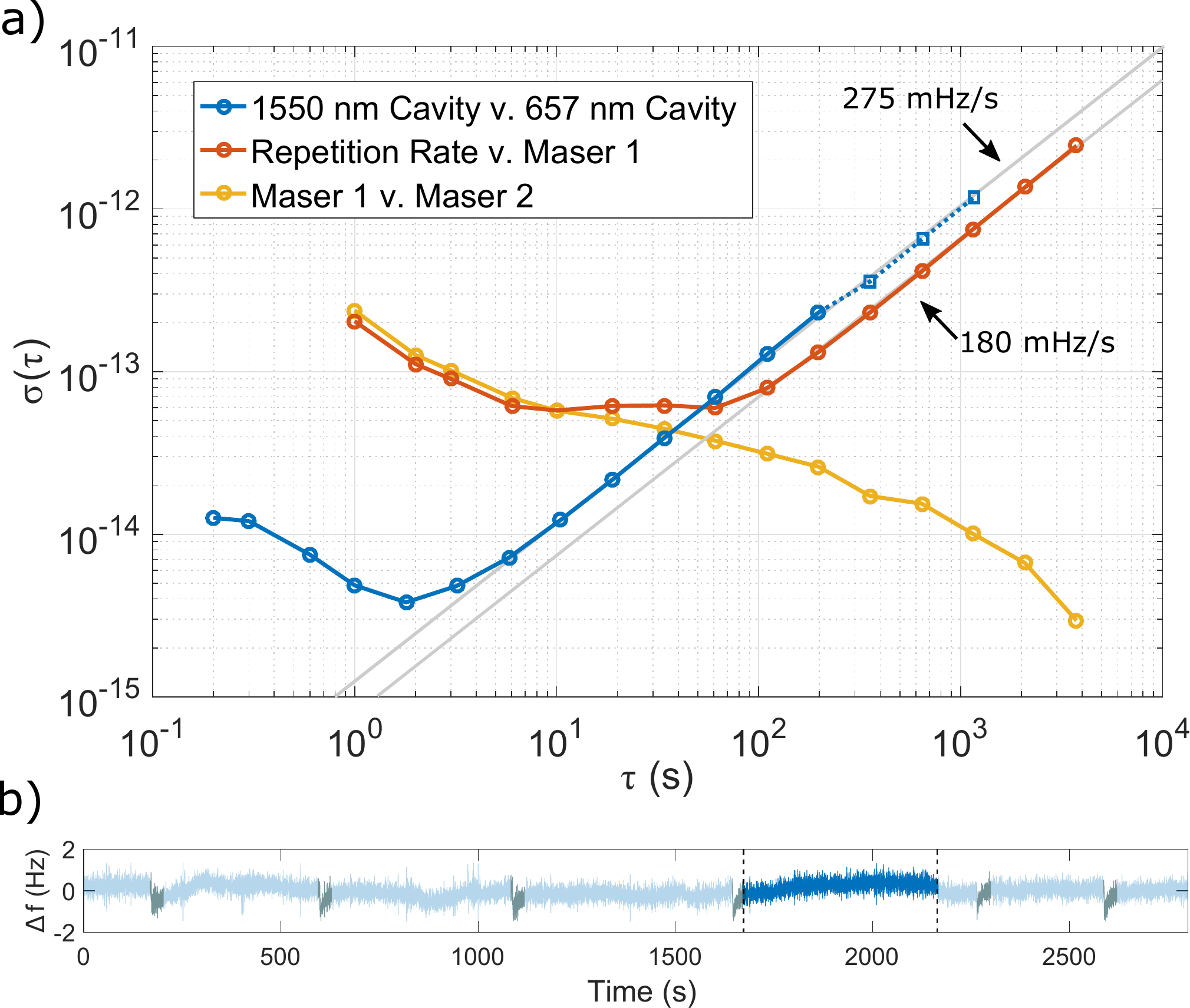}
  \caption{Allan deviation showing the relative stability of the two optical
    reference cavities (blue). At $\tau = 2$~s, the relative instability reaches
    a minimum value of $3.8\times10^{-15}$ while for long averaging times the
    relative cavity drift dominates and is determined to be 275~mHz/s. The red
    curve shows the comb repetition rate counted against a hydrogen maser while
    the comb is locked to beat $f'_a$. The maser stability alone (yellow) limits
    the observed Allan deviation for short time scales but the drift of the
    1550~nm reference cavity becomes apparent for $\tau > 100$~s. b) Counter
    record for the Allan deviation of the relative cavity instability (blue
    curve in a)) with linear drift removed.  Due to ambient noise amongst the
    laboratories involved in these measurements, glitches lasting several
    seconds appear sporadically in the counter record that are not readily
    detected in real-time operation (gray regions). As a result, the short
    timescale points in the Allan deviation are obtained from the highlighted
    500~s segment of the complete counter trace.  This portion represents
    low-noise perturbation-free system operation.}
  \label{fig:adev}
\end{figure}

Following $f_0$ subtraction, the comb is optically phase-locked to the
offset-free 657~nm beat $f'_b$ and, in doing so, the stability of the calcium
reference cavity is transferred across the entire comb bandwidth.  Using a
frequency counter (K+K FX80*) to record the out-of-loop offset-free beat $f'_a$
at 1550~nm yields the relative frequency stability of the two reference cavities
after scaling by the optical frequency of 193~THz.  The resulting Allan
deviation, displayed as the blue curve in Fig.~\ref{fig:adev}, shows both the
minimum relative instability of $3.8\times10^{-15}$ at $\tau = 2$~s and the
long-term relative cavity drift of 275~mHz/s.  This result is consistent with
the expected individual stability (1-3$\times10^{-15}$ at $\tau = 1$~s) of the
two cavity-stabilized lasers and thus there is no indication that our
waveguide-generated SC is introducing additional noise that limits the
measurement.


As a comparison and consistency check, the absolute drift rate of the 1550~nm
cavity alone is measured by locking the comb repetition rate to the offset-free
1550~nm beat $f'_a$ while simultaneously counting the repetition rate $f_r$
against a hydrogen maser (red curve in Fig.~\ref{fig:adev}).  Because $f'_a$ in
Eq.~\ref{eq:fa} is maser-referenced, long-term drifts in $f_r$ can only arise
from shifts in the reference cavity frequency $\nu_{1550}$. For short averaging
times ($\tau < 10$~s), the fractional stability obtained from the measurement
is limited by the maser reference (yellow curve in Fig.~\ref{fig:adev}), as
expected, to approximately $2\times10^{-13}/\sqrt\tau$. However, for
$\tau > 100$~s, the 180~mHz/s linear drift rate of the cavity becomes
apparent.

\section{Conclusion}
A key strength of these waveguides with respect to precision metrology
experiments is the broadly tunable design characteristics.  Because the SiN
devices presented here can be chosen to target narrow wavelength regions across
the visible and near-infrared, there is a clear path towards the all-in-one
frequency comparison chip presented in Fig.~\ref{fig:superchip} that can
generate all current optical clock wavelengths from a 1550~nm source.
Furthermore, the CMOS compatibility of SiN will allow integration with
photodection and feedback electronics for laser stabilization in an extremely
small and portable package.  Nevertheless, realizing such a chip will first
require improvements to the coupling loss between the input fiber and the
waveguide.  Fortunately, several different techniques have already been
demonstrated to improve the overall efficiency and, consequently, to reduce the
input laser power requirements~\cite{tiecke_efficient_2015,
  groblacher_highly_2013, cohen_optical_2013, shoji_low_2002,
  chen_low-loss_2010}.  Finally, we note that these SiN waveguide devices
should, in principle, be able to support higher precision measurements than
demonstrated here. However, future work will be needed to understand their
fundamental noise limitations in order to support the next-generation optical
clocks having stabilities approaching $1\times10^{-17}$ at $\tau = 1$~s.

\section*{Funding Information}
This research is supported by the Air Force Office of Scientific Research
(AFOSR) under award number FA9550-16-1-0016, the Defense Advanced Research
Projects Agency (DARPA) PULSE program, the National Aeronautics and Space
Administration (NASA), the National Institute of Standards and Technology
(NIST), and the National Research Council (NRC).

\section*{Acknowledgments}

The authors thank Wei Zhang and Frank Quinlan for helpful discussions on
frequency counting.

* Certain commercial equipment, instruments, or materials are identified in this
paper in order to specify the experimental procedure adequately. Such
identification is not intended to imply recommendation or endorsement by the
National Institute of Standards and Technology, nor is it intended to imply that
the materials or equipment identified are necessarily the best available for the
purpose.

This work is a contribution of the U.S. government and is not subject to
copyright.

See Supplement 1 for supporting content.

\bibliography{calciumcounting}

\begin{thebibliography}{10}
\newcommand{\enquote}[1]{``#1''}

\bibitem{halir_ultrabroadband_2012}
R.~Halir, Y.~Okawachi, J.~S. Levy, M.~A. Foster, M.~Lipson, and A.~L. Gaeta,
  \enquote{Ultrabroadband supercontinuum generation in a {{CMOS}}-compatible
  platform,} Optics Letters \textbf{37}, 1685--1687 (2012).

\bibitem{chavez_boggio_dispersion_2014}
J.~M. Chavez~Boggio, D.~Bodenm{\"u}ller, T.~Fremberg, R.~Haynes, M.~M. Roth,
  R.~Eisermann, M.~Lisker, L.~Zimmermann, and M.~B{\"o}hm, \enquote{Dispersion
  engineered silicon nitride waveguides by geometrical and refractive-index
  optimization,} Journal of the Optical Society of America B \textbf{31}, 2846
  (2014).

\bibitem{zhao_visible--near-infrared_2015}
H.~Zhao, B.~Kuyken, S.~Clemmen, F.~Leo, A.~Subramanian, A.~Dhakal, P.~Helin,
  S.~Severi, E.~Brainis, G.~Roelkens, and R.~Baets,
  \enquote{Visible-to-near-infrared octave spanning supercontinuum generation
  in a silicon nitride waveguide,} Optics Letters \textbf{40}, 2177 (2015).

\bibitem{epping_-chip_2015}
J.~P. Epping, T.~Hellwig, M.~Hoekman, R.~Mateman, A.~Leinse, R.~G. Heideman,
  A.~{van Rees}, P.~J. {van der Slot}, C.~J. Lee, C.~Fallnich, and K.-J.
  Boller, \enquote{On-chip visible-to-infrared supercontinuum generation with
  more than 495 {{THz}} spectral bandwidth,} Optics Express \textbf{23}, 19596
  (2015).

\bibitem{hsieh_supercontinuum_2007}
I.-W. Hsieh, X.~Chen, X.~Liu, J.~I. Dadap, N.~C. Panoiu, C.-Y. Chou, F.~Xia,
  W.~M. Green, Y.~A. Vlasov, and R.~M. Osgood, \enquote{Supercontinuum
  generation in silicon photonic wires,} Optics Express \textbf{15},
  15242--15249 (2007).

\bibitem{kuyken_octave-spanning_2015}
B.~Kuyken, T.~Ideguchi, S.~Holzner, M.~Yan, T.~W. H{\"a}nsch,
  J.~Van~Campenhout, P.~Verheyen, S.~Coen, F.~Leo, R.~Baets, G.~Roelkens, and
  N.~Picqu{\'e}, \enquote{An octave-spanning mid-infrared frequency comb
  generated in a silicon nanophotonic wire waveguide,} Nature Communications
  \textbf{6}, 6310 (2015).

\bibitem{oh_supercontinuum_2014}
D.~Y. Oh, D.~Sell, H.~Lee, K.~Y. Yang, S.~A. Diddams, and K.~J. Vahala,
  \enquote{Supercontinuum generation in an on-chip silica waveguide,} Optics
  Letters \textbf{39}, 1046 (2014).

\bibitem{oh_coherent_2017}
D.~Y. Oh, K.~Y. Yang, C.~Fredrick, G.~Ycas, S.~A. Diddams, and K.~J. Vahala,
  \enquote{Coherent ultra-violet to near-infrared generation in silica ridge
  waveguides,} Nature Communications \textbf{8}, 13922 (2017).

\bibitem{eggleton_chalcogenide_2011}
B.~J. Eggleton, B.~Luther-Davies, and K.~Richardson, \enquote{Chalcogenide
  photonics,} Nature Photonics \textbf{5}, 141--148 (2011).

\bibitem{dolgaleva_broadband_2010}
K.~Dolgaleva, W.~C. Ng, L.~Qian, J.~S. Aitchison, M.~C. Camasta, and M.~Sorel,
  \enquote{Broadband self-phase modulation, cross-phase modulation, and
  four-wave mixing in 9-mm-long {{AlGaAs}} waveguides,} Optics Letters
  \textbf{35}, 4093--4095 (2010).

\bibitem{johnson_octave-spanning_2015}
A.~R. Johnson, A.~S. Mayer, A.~Klenner, K.~Luke, E.~S. Lamb, M.~R.~E. Lamont,
  C.~Joshi, Y.~Okawachi, F.~W. Wise, M.~Lipson, U.~Keller, and A.~L. Gaeta,
  \enquote{Octave-spanning coherent supercontinuum generation in a silicon
  nitride waveguide,} Optics Letters \textbf{40}, 5117 (2015).

\bibitem{kippenberg_microresonator-based_2011}
T.~J. Kippenberg, R.~Holzwarth, and S.~A. Diddams,
  \enquote{Microresonator-{{Based Optical Frequency Combs}},} Science
  \textbf{332}, 555--559 (2011). 00601.

\bibitem{herr_temporal_2013}
T.~Herr, V.~Brasch, J.~D. Jost, C.~Y. Wang, N.~M. Kondratiev, M.~L. Gorodetsky,
  and T.~J. Kippenberg, \enquote{Temporal solitons in optical microresonators,}
  Nature Photonics \textbf{8}, 145--152 (2013).

\bibitem{kobayashi_optical_1988}
T.~Kobayashi, H.~Yao, K.~Amano, Y.~Fukushima, A.~Morimoto, and T.~Sueta,
  \enquote{Optical pulse compression using high-frequency electrooptic phase
  modulation,} IEEE Journal of Quantum Electronics \textbf{24}, 382--387
  (1988).

\bibitem{ishizawa_octave-spanning_2010}
A.~Ishizawa, T.~Nishikawa, A.~Mizutori, H.~Takara, S.~Aozasa, A.~Mori,
  H.~Nakano, A.~Takada, and M.~Koga, \enquote{Octave-spanning frequency comb
  generated by 250 fs pulse train emitted from 25 {{GHz}} externally
  phase-modulated laser diode for carrier-envelope-offset-locking,} Electronics
  Letters \textbf{46}, 1343--1344 (2010).

\bibitem{supradeepa_bandwidth_2012}
V.~R. Supradeepa and A.~M. Weiner, \enquote{Bandwidth scaling and spectral
  flatness enhancement of optical frequency combs from phase-modulated
  continuous-wave lasers using cascaded four-wave mixing,} Optics Letters
  \textbf{37}, 3066--3068 (2012).

\bibitem{bartels_10-ghz_2009}
A.~Bartels, D.~Heinecke, and S.~A. Diddams, \enquote{10-{{GHz
  Self}}-{{Referenced Optical Frequency Comb}},} Science \textbf{326}, 681--681
  (2009).

\bibitem{mayer_frequency_2015}
A.~S. Mayer, A.~Klenner, A.~R. Johnson, K.~Luke, M.~R.~E. Lamont, Y.~Okawachi,
  M.~Lipson, A.~L. Gaeta, and U.~Keller, \enquote{Frequency comb offset
  detection using supercontinuum generation in silicon nitride waveguides,}
  Optics Express \textbf{23}, 15440 (2015).

\bibitem{klenner_gigahertz_2016}
A.~Klenner, A.~S. Mayer, A.~R. Johnson, K.~Luke, M.~R.~E. Lamont, Y.~Okawachi,
  M.~Lipson, A.~L. Gaeta, and U.~Keller, \enquote{Gigahertz frequency comb
  offset stabilization based on supercontinuum generation in silicon nitride
  waveguides,} Optics Express \textbf{24}, 11043 (2016).

\bibitem{chang_heterogeneous_2017}
L.~Chang, M.~H.~P. Pfeiffer, N.~Volet, M.~Zervas, J.~D. Peters, C.~L.
  Manganelli, E.~J. Stanton, Y.~Li, T.~J. Kippenberg, and J.~E. Bowers,
  \enquote{Heterogeneous integration of lithium niobate and silicon nitride
  waveguides for wafer-scale photonic integrated circuits on silicon,} Accepted
  for publication in Optics Letters  (2017).

\bibitem{fortier_octave-spanning_2006}
T.~M. Fortier, A.~Bartels, and S.~A. Diddams, \enquote{Octave-spanning
  {{Ti}}:sapphire laser with a repetition rate $>$ 1 {{GHz}} for optical
  frequency measurements and comparisons,} Optics Letters \textbf{31},
  1011--1013 (2006).

\bibitem{nakajima_multi-branch_2010}
Y.~Nakajima, H.~Inaba, K.~Hosaka, K.~Minoshima, A.~Onae, M.~Yasuda, T.~Kohno,
  S.~Kawato, T.~Kobayashi, T.~Katsuyama, and {others}, \enquote{A multi-branch,
  fiber-based frequency comb with millihertz-level relative linewidths using an
  intra-cavity electro-optic modulator,} Optics Express \textbf{18}, 1667--1676
  (2010).

\bibitem{hagemann_providing_2013}
C.~Hagemann, C.~Grebing, T.~Kessler, S.~Falke, N.~Lemke, C.~Lisdat, H.~Schnatz,
  F.~Riehle, and U.~Sterr, \enquote{Providing $10^{-16}$ {{Short}}-{{Term
  Stability}} of a 1.5-\textmu{}m {{Laser}} to {{Optical Clocks}},} IEEE
  Transactions on Instrumentation and Measurement \textbf{62}, 1556--1562
  (2013).

\bibitem{fox_high_2012}
R.~W. Fox, J.~A. Sherman, W.~Douglas, J.~B. Olson, A.~D. Ludlow, and C.~W.
  Oates, \enquote{A high stability optical frequency reference based on thermal
  calcium atoms,} in \enquote{2012 {{IEEE International Frequency Control
  Symposium Proceedings}},}  ({IEEE}, 2012), pp. 1--3.

\bibitem{2015APS..DMP.D1071O}
J.~Olson, R.~Fox, E.~{de Carlos-Lopez}, C.~Oates, and A.~Ludlow, \enquote{Laser
  {{Frequency Stabilization Using}} a {{Calcium Ramsey}}-{{Bord{\'e}
  Interferometer}},} in \enquote{{{APS Division}} of {{Atomic}}, {{Molecular}}
  and {{Optical Physics Meeting Abstracts}},}  (2015).

\bibitem{ycas_pynlo:_2015}
G.~Ycas, D.~Maser, and D.~Hickstein, \enquote{{{PyNLO}}: {{Nonlinear Optics
  Modelling}} for {{Python}},}  (2015). Https://github.com/PyNLO.

\bibitem{hickstein_photonic-chip_2016}
D.~Hickstein, G.~Ycas, A.~Lind, D.~C. Cole, K.~Srinivasan, S.~Diddams, and
  S.~Papp, \enquote{Photonic-chip {{Waveguides}} for {{Supercontinuum
  Generation}} with {{Picojoule Pulses}},} in \enquote{Integrated {{Photonics
  Research}}, {{Silicon}} and {{Nanophotonics}},}  ({Optical Society of
  America}, 2016), pp. IM3A--2.

\bibitem{amorim_sub-two-cycle_2009}
A.~A. Amorim, M.~V. Tognetti, P.~Oliveira, J.~L. Silva, L.~M. Bernardo, F.~X.
  K{\"a}rtner, and H.~M. Crespo, \enquote{Sub-two-cycle pulses by soliton
  self-compression in highly nonlinear photonic crystal fibers,} Optics Letters
  \textbf{34}, 3851--3853 (2009).

\bibitem{fallahkhair_vector_2008}
A.~B. Fallahkhair, K.~S. Li, and T.~E. Murphy, \enquote{Vector {{Finite
  Difference Modesolver}} for {{Anisotropic Dielectric Waveguides}},} Journal
  of Lightwave Technology \textbf{26}, 1423--1431 (2008).

\bibitem{turner_tailored_2006}
A.~C. Turner, C.~Manolatou, B.~S. Schmidt, M.~Lipson, M.~A. Foster, J.~E.
  Sharping, and A.~L. Gaeta, \enquote{Tailored anomalous group-velocity
  dispersion in silicon channel waveguides,} Optics Express \textbf{14},
  4357--4362 (2006).

\bibitem{dudley_supercontinuum_2006}
J.~M. Dudley, G.~Genty, and S.~Coen, \enquote{Supercontinuum generation in
  photonic crystal fiber,} Reviews of Modern Physics \textbf{78}, 1135--1184
  (2006).

\bibitem{akhmediev_cherenkov_1995}
N.~Akhmediev and M.~Karlsson, \enquote{Cherenkov radiation emitted by solitons
  in optical fibers,} Physical Review A \textbf{51}, 2602 (1995).

\bibitem{hong_optical_2017}
F.-L. Hong, \enquote{Optical frequency standards for time and length
  applications,} Measurement Science and Technology \textbf{28}, 012002 (2017).

\bibitem{ma_delivering_1994}
L.-S. Ma, P.~Jungner, J.~Ye, and J.~L. Hall, \enquote{Delivering the same
  optical frequency at two places: Accurate cancellation of phase noise
  introduced by an optical fiber or other time-varying path,} Optics Letters
  \textbf{19}, 1777--1779 (1994).

\bibitem{hall_stabilizing_1999-1}
J.~L. Hall, M.~S. Taubman, S.~A. Diddams, B.~Tiemann, J.~Ye, L.-S. Ma, D.~J.
  Jones, and S.~T. Cundiff, \enquote{Stabilizing and measuring optical
  frequencies,} in \enquote{Proceedings of the {{International Conference}} on
  {{Laser Spectroscopy}},}  ({World Scientific}, 1999).

\bibitem{stenger_ultraprecise_2002}
J.~Stenger, H.~Schnatz, C.~Tamm, and H.~R. Telle, \enquote{Ultraprecise
  {{Measurement}} of {{Optical Frequency Ratios}},} Physical Review Letters
  \textbf{88}, 07361 (2002).

\bibitem{tiecke_efficient_2015}
T.~G. Tiecke, K.~P. Nayak, J.~D. Thompson, T.~Peyronel, N.~P. de~Leon,
  V.~Vuleti{\'c}, and M.~D. Lukin, \enquote{Efficient fiber-optical interface
  for nanophotonic devices,} Optica \textbf{2}, 70--75 (2015).

\bibitem{groblacher_highly_2013}
S.~Gr{\"o}blacher, J.~T. Hill, A.~H. Safavi-Naeini, J.~Chan, and O.~Painter,
  \enquote{Highly efficient coupling from an optical fiber to a nanoscale
  silicon optomechanical cavity,} Applied Physics Letters \textbf{103}, 181104
  (2013).

\bibitem{cohen_optical_2013}
J.~D. Cohen, S.~M. Meenehan, and O.~Painter, \enquote{Optical coupling to
  nanoscale optomechanical cavities for near quantum-limited motion
  transduction,} Optics Express \textbf{21}, 11227 (2013).

\bibitem{shoji_low_2002}
T.~Shoji, T.~Tsuchizawa, T.~Watanabe, K.~Yamada, and H.~Morita, \enquote{Low
  loss mode size converter from 0.3 \textmu{}m square {{Si}} wire waveguides to
  singlemode fibres,} Electronics Letters \textbf{38}, 1669--1670 (2002).

\bibitem{chen_low-loss_2010}
L.~Chen, C.~R. Doerr, Chen, and T.-Y. Liow, \enquote{Low-{{Loss}} and
  {{Broadband Cantilever Couplers Between Standard Cleaved Fibers}} and
  {{High}}-{{Index}}-{{Contrast Si}}$_{3}${{N}}$_{4}$ or {{Si Waveguides}},}
  IEEE Photonics Technology Letters \textbf{22}, 1744--1746 (2010).

\end{thebibliography}


\pagebreak
\section*{Supplementary Material}
\begin{figure}[b!]
  \centering
  \includegraphics[width=0.7\textwidth]{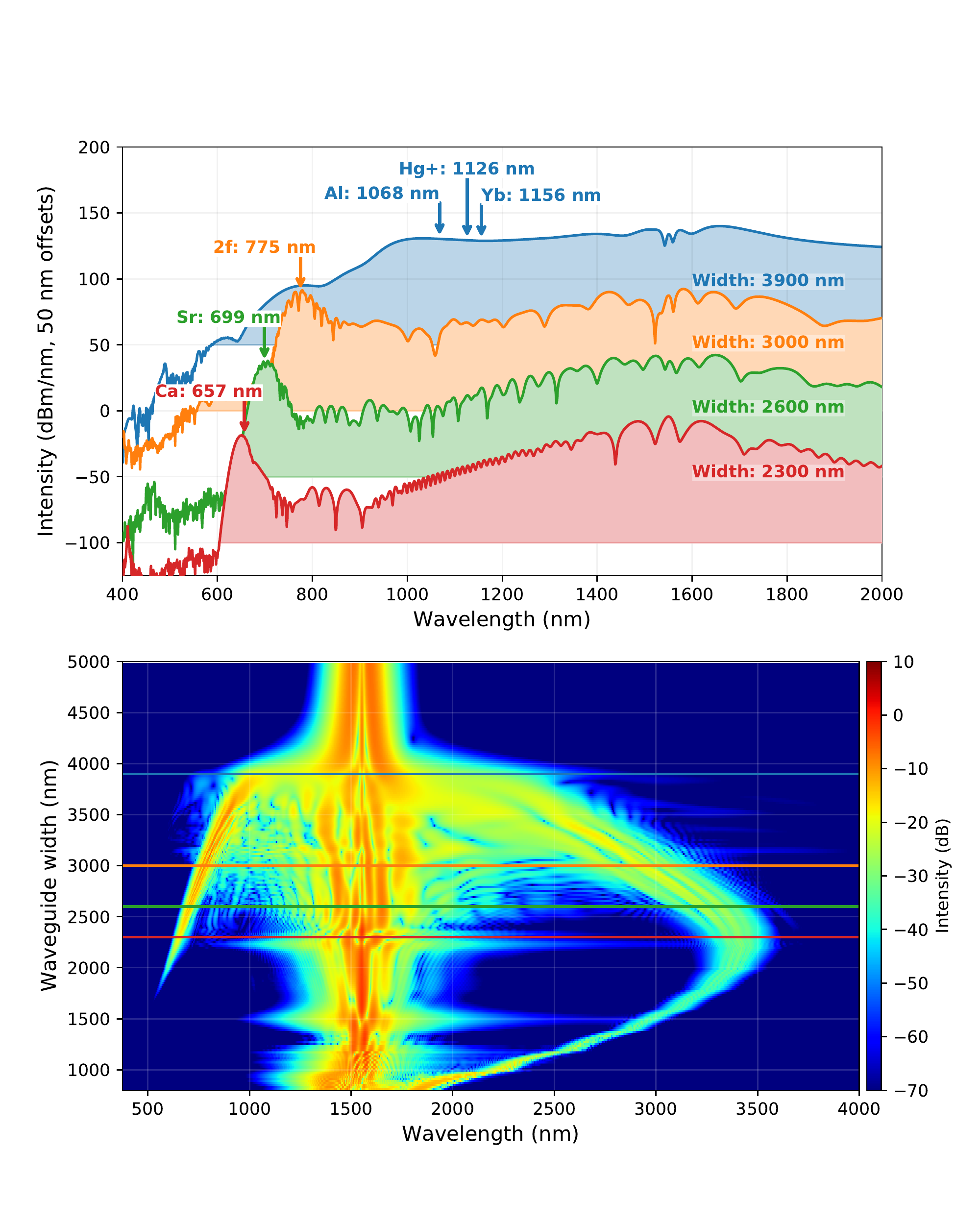}
  \caption{a) Simulated spectra for 20~mm long air-clad SiN waveguides with a
    thickness of 700~nm. A single chip with this geometry can simultaneously
    cover all standard optical clock wavelengths with sufficient intensity for
    optical heterodyne beats (specific clock transitions denoted with
    arrows). b) Supercontinuum spectra for waveguide widths swept continuously
    from 800~nm to 5000~nm.  Colored horizontal lines indicate the locations of
    the line-out spectra shown in a).  }
  \label{fig:spectrum_suppl}
\end{figure}

The availability of air-clad 600~nm thickness silicon nitride (SiN) with oxide
undercladding at the time of the experiment led us to use this chip geometry
since it provides a suitable dispersion profile for visible dispersive wave
generation with 1550~nm pumping.  However, to also achieve sufficient optical
power for heterodyne beats in the 1100~nm wavelength range where sub-harmonics
of other optical clocks like Yb, Al$^+$, and Hg$^+$ are typically accessed, a
thicker SiN layer should ideally be used instead. Fig.~\ref{fig:spectrum_suppl}
shows simulated supercontinuum (SC) spectra for 700~nm thickness air-clad SiN
using the same pulse parameters as were used to generate Fig.~2a in the main
text.  In order to enable integration of all clock-wavelength-generating
waveguides on the same chip, it is important, from a fabrication perspective,
that they can all be created with a single uniform-thickness nitride layer.

\end{document}